\documentclass[pre,twocolumn,aps,superscriptaddress,showpacs]{revtex4-1}

\usepackage{graphicx}
\usepackage{amssymb}
\usepackage{mathrsfs}
\usepackage[english]{babel}
\usepackage{amstext}
\usepackage{amsmath}
\usepackage{bm}
\usepackage{xcolor}
\usepackage{subfig}

{

}
\graphicspath{{./}{../}}

\begin{document}
\title{\Large Atomic-scale expressions for viscosity and fragile-strong behavior in metal alloys based on the Zwanzig-Mountain formula}

\author{G. Chevallard}
\affiliation{Department of Physics ``A. Pontremoli'', University of Milan, via Celoria 16, 20133 Milan, Italy}

\author{K. Samwer}
\affiliation{I. Physikalisches Institut, University of Goettingen, Germany.}

\author{A. Zaccone}
\affiliation{Department of Physics ``A. Pontremoli'', University of Milan, via Celoria 16, 20133 Milan, Italy}
\affiliation{Department of Chemical Engineering and Biotechnology, University of Cambridge, Cambridge
CB3 0AS, U.K.}

\begin{abstract}
We combine the shoving model  of $T$-dependent viscosity of supercooled liquids with the Zwanzig-Mountain formula for the high-frequency shear modulus, using the $g(r)$ of MD simulations of metal alloys as the input. This scheme leads to a semi-analytical expression for the viscosity as a function of temperature, which provides a three-parameter model fitting of experimental data of viscosity for the same alloy for which $g(r)$  was calculated. The model provides direct access to the influence of atomic-scale physical quantities such as the interatomic potential $\phi(r)$, on the viscosity and fragile-strong behavior. In particular, it is established that a steeper interatomic repulsion leads to fragile liquids, or, conversely, that "soft atoms make strong liquids". 

\end{abstract}


\maketitle

\section{Introduction}
Different views of the glass transition have led to quite different descriptions of the viscosity of supercooled liquids. The kinetic view of the glass transition, which relies on a substantial continuity between liquid and solid glass, goes back to pioneering ideas of Y. Frenkel~\cite{Frenkel,Trachenko} and provides the basis for Dyre's shoving model and its ramifications.
The entropic view of the glass transition, instead, based on the Adam-Gibbs scenario and later developed into a random first-order (ideal glass) transition, has led to a suitably modified Vogel-Fulcher-Tammann (VFT) equation with parameters that can be related to the entropy of cooperatively rearranging zones~\cite{Thirumalai}. Another approach based on the Adam-Gibbs scenario led to the Mauro equation for the viscosity~\cite{Mauro}. Yet a different type of approach based on Doremus' model~\cite{Doremus} of viscosity where bonds are broken under shear flow leads to a two-exponential form for the viscosity as a function of temperature~\cite{Ojovan}, which typically provides a better fitting to experimental data compared to single-exponential expressions~\cite{BulletinMRS}.

In general, closed-form expressions for the viscosity of supercooled liquids contain three fitting parameters, which are typically related to microscopically poorly defined quantities such as free volume or entropy. 
The shoving model developed by Dyre provides a different approach in this sense, as it links viscosity to thermally-activated jumps of atoms out of the nearest-neighbor cage, as in Frenkel's and Eyring's early approaches, with an activation energy which is described rigorously by means of continuum mechanics. 
In particular, the activation energy is expressed via the product of the high-frequency shear modulus of the liquid and an activation volume, which  follows from the analysis of the work done by a particle to shove around the surrounding atoms to escape from the cage. A similar relation between energy barrier and shear modulus is provided by the Cooperative Shear Model (CSM) where, however, the characteristic volume is larger and can be connected to the concept of shear transformation zones (STZs)~\cite{CSM}.

The shoving model provides the starting point for a more microscopic description of the viscosity and relaxation time of supercooled liquids. In particular, upon approximating the shear modulus $G_\infty$ with Born-Huang (affine) lattice dynamics (as appropriate for the high-frequency modulus), $G_\infty$ can be directly related to the short-range part of the radial distribution function (RDF) $g(r)$, and hence to the interatomic potential. This led to the Krausser-Samwer-Zaccone (KSZ) equation~\cite{krausser2015interatomic}, which expresses the $T$-dependent viscosity in closed-form in terms of the thermal expansion coefficient $\alpha_{T}$, the interatomic repulsion steepness parameter $\lambda$ (obtained from a power-law fitting of the RDF up to the maximum of the first peak) and the activation volme $V_{c}$ mentioned above. The KSZ equation reads as
\begin{equation}\label{thvisc}
\frac{\eta(T)}{\eta_{0}}
=
\exp{\left\{ \frac{V_c C_G}{k_{B}T}		\exp{\left[		(2+\lambda)	\alpha_T T_g \left(	1-\frac{T}{T_g}\right)	\right]}		\right\}}, \nonumber
\end{equation}
where $C_G$ is the value of the $G_\infty$ at $T_g$, again evaluated analytically with the Born-Huang formula. This equation provides a two-parameter fitting of viscosity data since $\lambda$ is determined by fitting of the $g(r)$ data, $\alpha_{T}$ is an experimentally determined quantity, which leaves $\eta_0$, $T_g$ and $V_c$  as the only parameters, with the important constraint that $V_c \sim 10^{-30}$m$^{3}$. The glass transition temperature $T_g$, in the above formula, defines to the temperature at which the glass state loses its rigidity~\cite{zaccone2013disorder}.
Furthermore, it has been shown that the repulsion parameter $\lambda$ can be mapped accurately on realistic microscopic parameters of Thomas-Fermi screening length and Born-Mayer repulsion for the electron-gas screened repulsion between two ions in metals. 

In spite of its success in providing, for the first time, a direct connection between viscosity (and fragility) and the microscopic physics of atomic-scale structure and interactions~\cite{lagogianni2016unifying}, the KSZ equation suffers from an intrinsic ambiguity in determining the $\lambda$ parameter from experimental or simulated $g(r)$ data. While the fitting protocol of Ref.~\cite{krausser2015interatomic} provides a consistent assessment of the effect of interatomic repulsion on $T$-dependent viscosity and the fragility $m$ (defined as slope of $\eta(T)$ near $T_{g}$), other protocols~\cite{Kelton} have provided ambiguous results. In particular, the original protocol of Ref.~\cite{krausser2015interatomic} has shown that a steeper interatomic repulsion results in a fragile behavior of the glass-forming liquid, whereas a softer repulsion is associated with strong liquids. 
A different protocol for extracting $\lambda$ with less prescriptive constraints on the fitting was used in Ref.~\cite{Kelton}. In that approach, $\lambda$ was taken to be a free parameter which also depends on temperature, and the opposite scenario, i.e. softer repulsion leads to fragile behavior, was found. However, it has been later demonstrated that $\lambda$ does not depend on temperature~\cite{Wang}, which invalidates this fitting protocol. 

Here we develop a different, perhaps more sophisticated, approach, which combines the shoving model with the microscopic Zwanzig-Mountain formula for the $G_\infty$ of liquids. This leads to semi-analytical expressions for $\eta(T)$ and for $m$, which directly link these quantities to the $g(r)$ and to the interatomic potential $\phi(r)$. Upon successfully calibrating these expressions for the case of Cu$_{50}$Zr$_{50}$, a new interatomic repulsion paramter $l$ is identified which is unambiguously linked with the repulsive part of $\phi(r)$. Upon letting this parameter vary, fictive materials with different interatomic repulsion softness are generated. The model analysis demonstrates that steeper interatomic repulsion leads to fragile behavior, thus reaffirming the correctness of the fitting protocol of Ref.~\cite{krausser2015interatomic} for the identification of $\lambda$ in the KSZ equation above. It also confirms the qualitative increasing trend of fragility $m$ increasing with potential repulsion steepness $l$ or $\lambda$, and recovers the linear trend already seen for $m(\lambda)$ in Ref.~\cite{krausser2015interatomic}.

\section{Theory}
\subsection{The shoving model}
We base our derivation of the viscosity of liquid metals here upon the so-called shoving model  \cite{dyre1998source}. The assumption at the basis of this model is that, within the transition state theory~\cite{Frenkel}, the activation energy of the average relaxation time is determined by the work done in shoving aside the surrounding liquid to allow "flow events". The main result of this model is the temperature dependence of the viscosity, which is related to the temperature dependence of the high frequency limit of the shear modulus $G_{\infty}(T)$:
\begin{equation}\label{shovingmodel}
\eta(T)=\eta_0\exp\left[\frac{V_c G_{\infty}(T)}{k_BT}\right]
\end{equation}
where $\eta_0$ is a constant prefactor and $V_c$ is the characteristic volume of the group of atoms involved in the shoving event (on the same order of magnitude of the nearest-neighbour cage), and is a weakly dependent function of $T$ such that this dependence is typically neglected. 
The $T$-dependence of viscosity $\eta$ is thus directly controlled by the $T$-dependence of $G_\infty$, while $V_c$ has a dependence on temperature which is more difficult to assess and for simplicity is normally taken as $T$-independent~\cite{dyre1998source}. The values of $V_c$ typically found are on the order of the atomic size, which is much smaller than the values of the characteristic volume normally found within the CSM model. 
In Frenkel's original derivation, activation energy in the Arrhenius exponential of Eq. (1) is given by $E=8\pi G_{\infty} r \Delta r^2$, where $r$ is the cage radius and $\Delta r$ is the increase of the cage radius due to thermal fluctuations which enables the atom to escape the cage (see page 193 in Ref.\cite{Frenkel}). 

\subsection{High-frequency shear modulus from the Zwanzig-Mountain formula}
Zwanzig and Mountain derived a popular expression for the high-frequency shear modulus of liquids, which reads as~\cite{zwanzig1965high}:
\begin{equation}\label{shearzwanzwig}
G_\infty = \rho k_B T + \frac{2 \pi}{15} \rho^2 \int _0^{\infty} dr g(r) \frac{d}{dr} \left[ r^4 \frac{d\phi}{dr}\right],
\end{equation}
where $g(r)$ denotes the radial distribution function (RDF) of the atoms in the liquid (average of the atomic species in an alloy), $\rho (T)$ is the total atomic density, and $\phi(r)$ is the average interatomic potential between any two atoms. 

Two crucial input to evaluate $G_\infty$ are, therefore, the RDF $g(r)$ and the interatomic potential $\phi(r)$.
In lieu of a theoretical expression for $g(r)$ valid for real metal alloys, which is obviously beyond reach, we base our analysis on MD simulations data of the average $g(r)$ in Cu$_{50}$Zr$_{50}$ alloys, from Ref.\cite{lagogianni2016unifying}. 

In order to provide an analytical handle on the various features of the $g(r)$, which in turn are connected to features of the interatomic potential $\phi(r)$ as shown below, we proceed to the following parameterization of the $g(r)$.

\subsection{Analytical parameterization of $g(r)$}
The parameterization of $g(r)$ is given by a sum of three terms, and it depends on 6 parameters. 
The first term describes, at the same time, the short-range repulsion and impenetrability of the atoms as well as behavior for large interatomic separations. This is mathematically expressed by the following asymptotic limits: $g(r)$ goes to zero for small $r$ and tends to one for $r \rightarrow \infty$. These asymptotic behaviors are encoded in the hyperbolic tangent. 
The second term represents the first nearest-neighbor shell, that corresponds to the first peak of the RDF, while the third term describes the decrease of structure with the distance, that corresponds to the decreasing height of the second peak and of the following Friedel oscillations. 
The first peak is given by a Gaussian, while the second peak and the Friedel oscillations are described by a decaying exponential multiplied by an oscillating function. 
Here we report the parameterization:
\begin{align}\label{firstterm}
g(r)=&\tanh \left[\left(\frac{r}{a}\right)^l\right]+ k\exp \left[-\frac{(r-a)^2}{b}\right]\\ \nonumber
&+\tanh \left[\left(\frac{r}{a}\right)^l\right] \sin \left(\frac{wr}{a} \right) \exp \left( -\frac{r}{h} \right).
\end{align}
This expression contains six parameters for which we now give a qualitative description. Parameter $a$ is approximately the position of the first peak (i.e.  the center of the Gaussian). Parameter $b$ is linked to the width of the first peak since it's proportional to the variance of the Gaussian. Parameter $l$, which is the most important for our subsequent analysis, gives the asymptotic power-law trend of the first term, hence it represents the steepness of the ascending part of the first peak of $g(r)$.
Parameter $k$ is the height of the Gaussian distribution, so it is linked to the height of the peak, which is also influenced by the other two terms. Finally, $w$ is related to the frequency of the Friedel oscillations while $h$ controls the decay of the envelop of the oscillations. 

The fitting of the MD simulations data of $g(r)$ for the Cu$_{50}$Zr$_{50}$ system is shown in Fig. 1 (a), and has been obtained with the following values of parameters: $k = 1.646$ (dimensionless),
$a = 2.716$ (Angstrom), $b = 0.05491$ (Angstrom$^2$), $l = 18.77$ (dimensionless), 
$h = 4.589$ (Angstrom),  $w = 7.265$ (dimensionless).
It is also shown (inset of Fig. 1 (a)) that the ascending part of $g(r)$, before the first peak, is perfectly described by a power-law $\sim(r/a)^{l}$, with $l=20$, over more than two decades in $r$.

\begin{figure}[b]
    \centering
    \includegraphics[width=0.85\columnwidth]{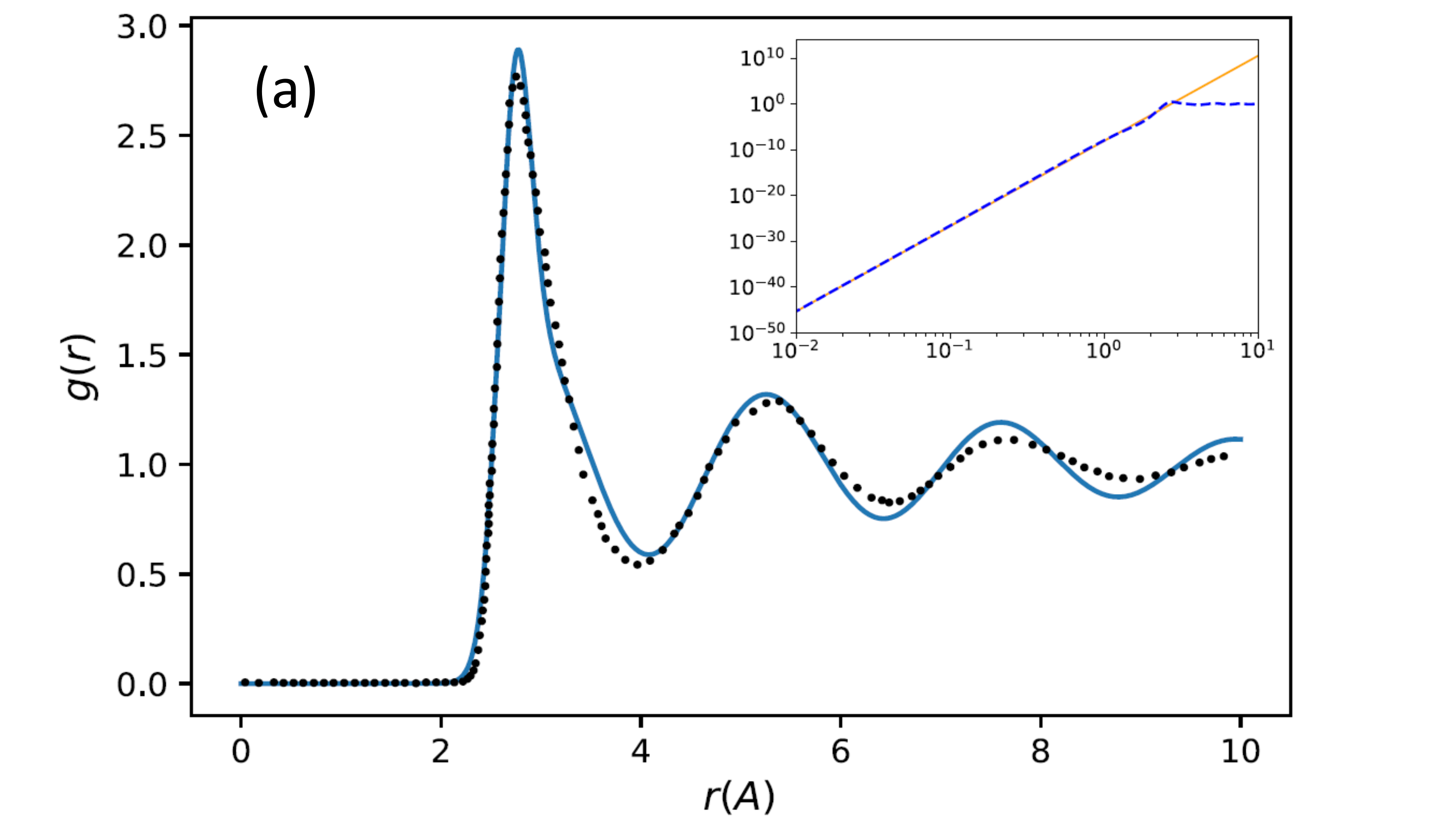}
    \vspace{0.2cm}
    \includegraphics[width=0.85\columnwidth]{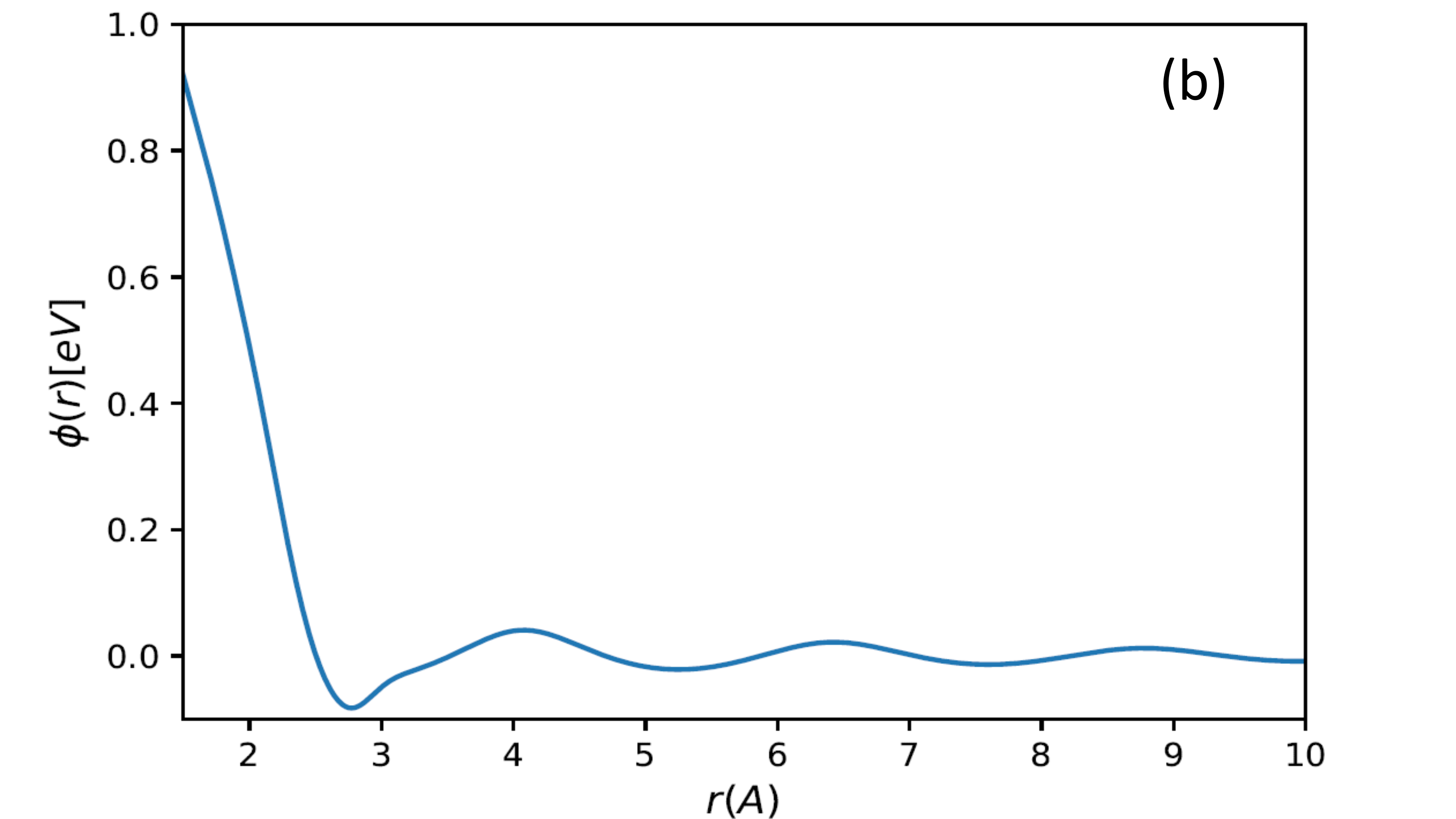}
    \caption{\textbf{(a)}: Analytical parameterization of the $g(r)$ simulations data for the binary Cu$_{50}$Zr$_{50}$ alloy. The values of the parameters are reported in the main text. The inset shows the analytical fit of the main panel together with a power-law trend line $\sim (r/a)^l$, with $l = 20$. \textbf{(b)}: The average interatomic potential (potential of mean force) obtained from the analytical fitting of $g(r)$.}
    \label{fig:tre}
\end{figure}

\subsection{Determination of the interatomic potential $\phi(r)$}
The RDF is related to the potential of mean force, $\phi(r)$, via the reversible work theorem:
\begin{equation}\label{meanforcepotential}
\beta \phi(r) = -\ln g(r)
\end{equation}
the proof of which can be found in the textbooks~\cite{chandler1987introduction}. 
The mean interatomic potential between two atoms $\phi(r)$ determined in this way thus accounts for many-body effects from the surrounding electronic and atomic environment. Using the fitting of the MD $g(r)$ data for the Cu$_{50}$Zr$_{50}$ alloy we obtain the interatomic potential profile shown in Fig. 1 (b). The potential features a rather steep interatomic repulsion due closed electron shell repulsion followed an attractive bonding minimum mediated by the nearly-free electrons. After the minimum, the oscillations represent the Friedel oscillations in the electronic density.

\begin{figure}
    \centering
    \includegraphics[width=0.9\linewidth]{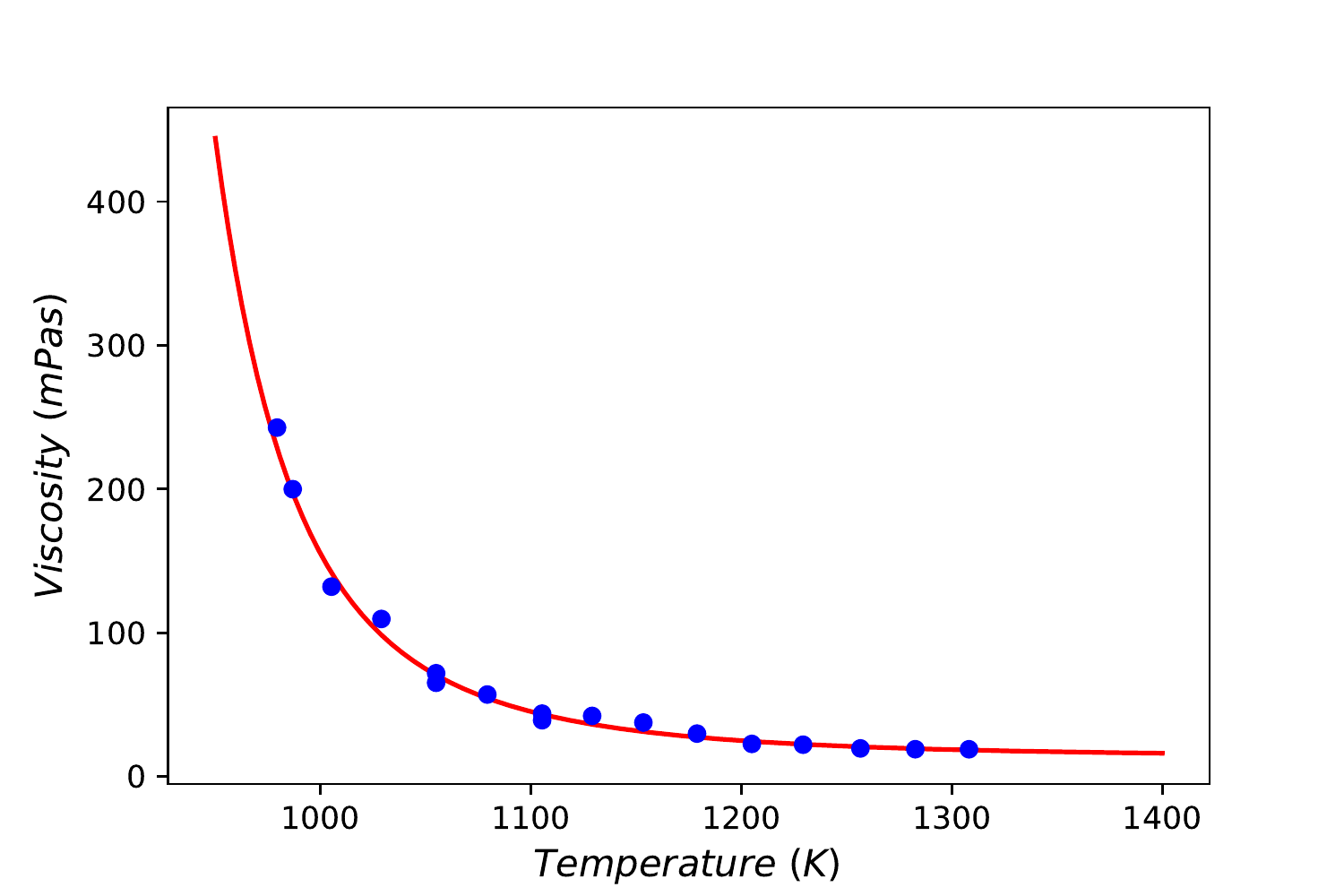}
    \caption{Model fitting of experimental viscosity data from Ref.~\cite{lagogianni2016unifying} using Eq.~\eqref{viscosity} with $V_c=5.6 \cdot 10^{-29}$m$^3$, $\eta_0=14$ mPa$\cdot$s, $I$ given by the integral in Eq.~\eqref{shearzwanzwig} with $\phi(r)$  given by Eq.~\eqref{meanforcepotential} and $g(r)$ from the analytical parameterization Eq.~\eqref{firstterm} of MD simulations data of $g(r)$ from Ref.~\cite{lagogianni2016unifying}. An effective value of thermal expansion coefficient has been taken, $\alpha_{T}=0.00377$K$^{-1}$ for the comparison.}
    \label{fig:uno}
\end{figure}

\section{Results and discussion}
\subsection{$T-$dependent viscosity}
Using Eqs. (3)-(4) calibrated on the $g(r)$ data for the Cu$_{50}$Zr$_{50}$ alloy inside the Zwanzig-Mountain formula Eq. (2), we are now able to evaluate the $T$-dependent viscosity $\eta$ by means of the shoving model, Eq. (1).

Upon denoting the $k_{B}T$-normalized integral in the Zwanzig-Mountain formula Eq. (2) as $I$, we therefore arrive at the following expression for the viscosity:
\begin{equation}\label{viscosity}
\frac{\eta}{\eta_0} = \exp\left[V_c \rho(T)\left(1+\frac{2\pi}{15} \rho(T)I(l)\right)\right].
\end{equation}
Note the cancellation of a $k_{B}T$ factor contained in $G_{\infty}$ (and recall that $\phi(r)=-k_{B}T \ln g(r)$) with the $k_{B}T$ factor in the denominator of the argument of the exponential in Eq. (1).
Here the $T$-dependent density $\rho$ is expressed in terms of the thermal expansion coefficient $\alpha_T$, using the definition of the latter $\alpha_T =V^{-1} (\partial V/\partial T)_T$, leading to:
\begin{equation}
\rho(T)=\rho_0 \exp \left[-\alpha_T \left(T-T_0\right)\right]
\end{equation}
where $\rho_0=7 \cdot 10^{3}$[kg/m$^3$] is a known value of density at the reference temperature $T_0=298$K~\cite{Baosheng}. The above relation is normally linearized due to the small value of $\alpha_{T}$, leading to 
$\rho(T)\approx \rho_0\left[1-\alpha_T \left(T-T_0\right)\right]$.
Also, one should note that the integral $I$ in Eq.~(2) is also $T$-dependent due to the $k_{B}T$  factor in the definition of $\phi(r)$ in Eq.~(4).

In Fig. 2 we report the fitting of the viscosity data of Cu$_{50}$Zr$_{50}$ as a function of temperature ( measured with a levitating drop method in Ref.~\cite{lagogianni2016unifying}) by means of Eq.~\eqref{viscosity}.
There are two fitting parameters in the comparison, one is the effective thermal expansion coefficient 
$\alpha_{T} = 0.00377$K$^{-1}$, which is significantly larger than the typical values for metallic melts ($\sim 10^{-4}$K$^{-1}$) and effectively compensates for the neglected $T$-dependence of the activation volume $V_c$ and of the atomic structure given by $g(r)$, since their $T$-dependence is not known. In particular, the $T$-dependence of $V_{c}$ may play an important role, since it was shown to be a rapidly decreasing function of $T$ upon approaching $T_{g}$ from below~\cite{Schwabe}. However, the dependence of $V_{c}$ on $T$ in the high temperature liquid phase is not known, and this may explain the larger value of the fitted $\alpha_{T}$ coefficient, which makes up for neglecting the decrease of $V_c$ upon increasing $T$.

The other fitting parameter is $V_c$ which is found to be equal to $5.6 \cdot 10^{-29}$m$^3$, i.e. in close agreement with typical values of the shoving volume found in previous works~\cite{krausser2015interatomic}, and corresponds to the characteristic size of the nearest-neighbor cage in disordered metals. However, it is significantly smaller than the typical size of a cooperative flow event~\cite{CSM}.

All in all, Eq.~(5) provides a three-parameter fit of viscosity data over a broad range of $T$, and, unlike other popular three-parameter models such as VFT, the Avramov-Milchev (AM) equation~\cite{Avramov} and the Mauro equation~\cite{Mauro}, all the parameters can be traced back to atomic-scale structure and interactions, in that being similar to the KSZ equation~\cite{krausser2015interatomic}. The latter still retains the favorable advantage, over Eq.~(5) of being in simple, fully analytical form. 

\subsection{Effect of interatomic potential on viscosity and fragile-strong behavior}
Thanks to the direct connection that the above model provides between $\eta(T)$ and microscopic atomic-scale parameters, it is possible to analyze the effect of the interatomic potential $\phi(r)$ on the viscosity and on the fragile-strong behavior. We use the interatomic repulsion steepness parameter $l$ in Eq.~(3) as a proxy to design fictive materials of varying interatomic repulsion, with the aim of studying the effect of the interatomic repulsion on viscosity and fragility. 

We start from the level of the RDF $g(r)$ and vary the repulsion steepness parameter $l$ around the value ($l =20$) that we found in the fitting of MD simulation data (Fig. 1 (b)). We thus obtain the fictive RDF's shown in Fig. 3 (a). It is clear that large values of $l$ correspond to a sharp rise of the first peak of the RDF, whereas low values of $l$ correspond to a less steep rise of the peak. Using Eq.~(4) we then obtain the corresponding potentials of mean force describing the interatomic potential, shown in Fig. 3 (b) for the same values of $l$ shown in Fig. 3 (a). It is evident that large $l$ values ($l \gtrsim 20$) result in steep interatomic repulsion, whereas low values of $l$ (i.e. $l < 10-20$) result in softer repulsive potentials. 
Based on Eq. (3), when $r/a < 1$ (i.e. at short range before the first peak of the RDF) the $\tanh$ can be approximated as a linear function, hence $\sim (r/a)^{l}$. Before the peak, the parameter $l$ is thus closely related to the $\lambda$ parameter of the KSZ equation, defined as $g(r) \sim (r-\sigma)^{\lambda}$ where $\sigma$ is a hard-core atomic size. Even though the power-law trend to describe the interatomic repulsion is a common feature of KSZ and of the present approach, the presence of a hard-core cut-off size $\sigma$ in the KSZ model is a significant difference, which quantitatively may lead to different values of $l$ and $\lambda$, although it should not affect the qualitative trends.

It is interesting to note that the parameter $l$ affects only the repulsive part of the potential, whereas, remarkably, the attractive part of the potential, from the minimum on, is almost unaffected. 
This allows us to single out the effect of the interatomic repulsion, which strongly depends on the atomic composition of the metal.

\begin{figure}[b]
    \centering
    \includegraphics[width=0.85\columnwidth]{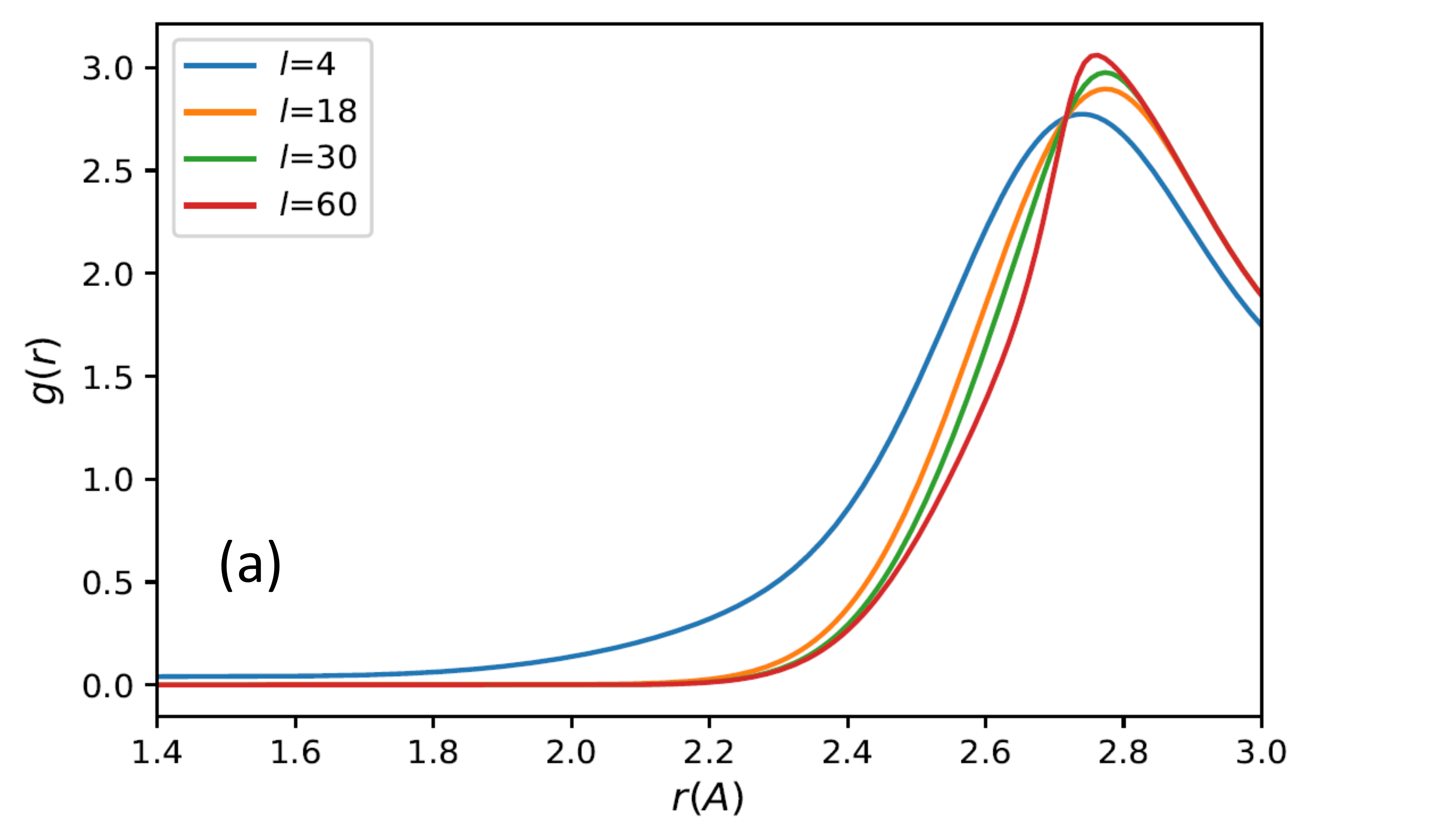}
    \vspace{0.2cm}
    \includegraphics[width=0.85\columnwidth]{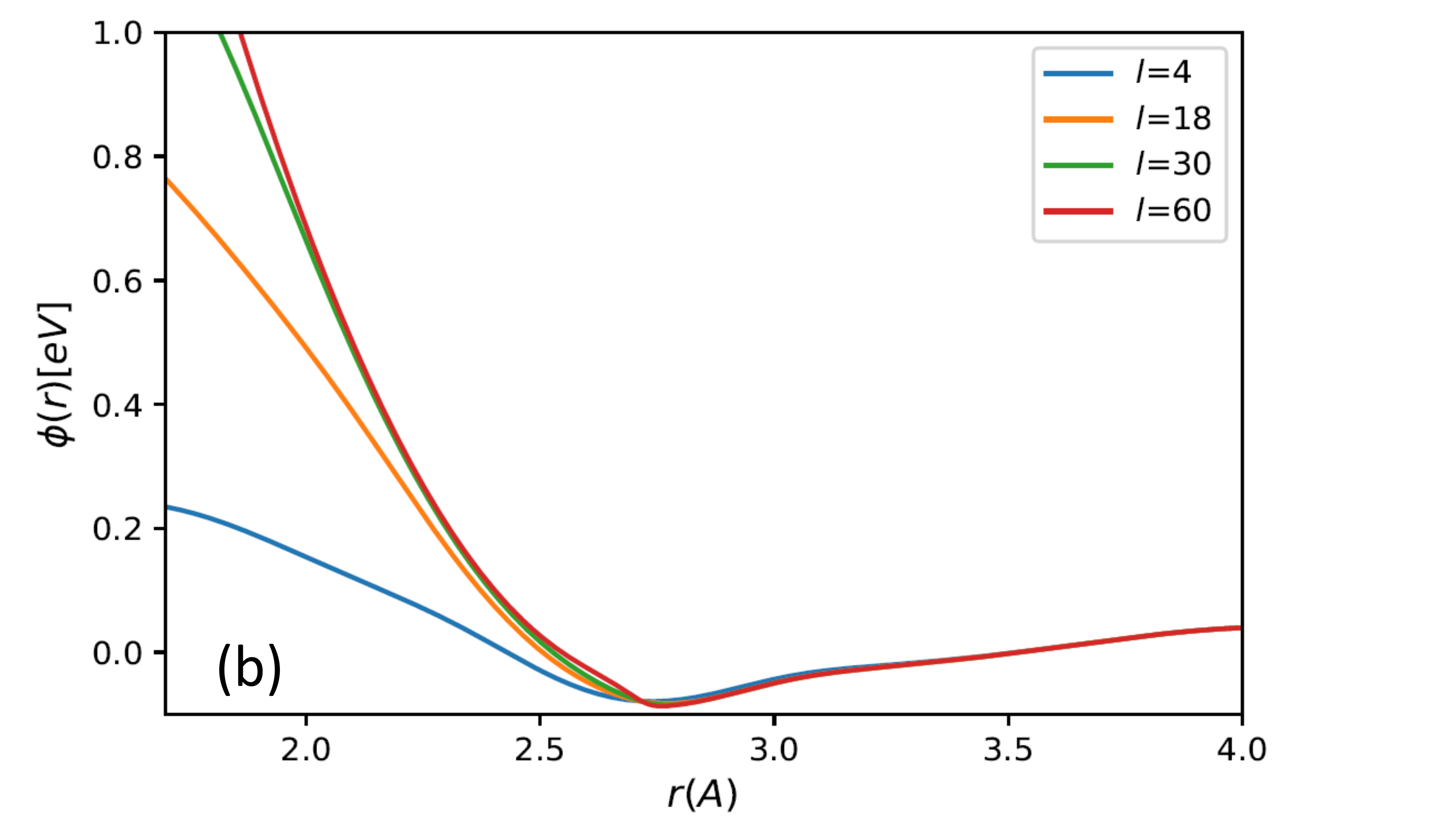}
    \caption{\textbf{(a)}: Evolution of the $g(r)$ upon varying the interatomic steepness parameter $l$ around the value $l\approx 20$ used in the fitting of the $g(r)$ in Fig. 1 (a). From left to right $l =4, 18, 30, 60$.\textbf{(b)}: The average interatomic potential (potential of mean force) for the same values of the $l$ parameter shown in panel (a).}
    \label{fig:tre}
\end{figure}

The corresponding viscosities as a function of $T$ calculated based on the RDFs and $\phi(r)$ profiles of Fig. 3 are shown in Fig. 4 (a). Upon varying the interatomic steepness $l$, it is clear that the slope of the $\eta(T)$ changes. In particular, larger values of steepness $l$ correspond to larger (steeper) slopes of the viscosity curves. The microscopic explanation for this behavior resides in the $k_{B}T$-normalized integral $I$ in the ZM formula Eq. (2). The value of $I$ in Eq. (4) increases systematically upon increasing $l$, which results in a $G_\infty$ more steeply rising with $T$, hence in a larger slope of $\eta(T)$. This is mainly due to the factor $\frac{d}{dr} \left[r^4 \frac{d\phi}{dr}\right]$ inside the integral, which clearly gives larger contributions as the repulsive decay of $\phi(r)$ becomes steeper. This fact can be easily checked on the simple example of a power-law repulsive decay $\phi(r) \sim r^{-n}$: the larger the exponent $n$ the larger the contribution of this factor to the integral. 

\begin{figure}[b]
    \centering
    \includegraphics[width=0.85\columnwidth]{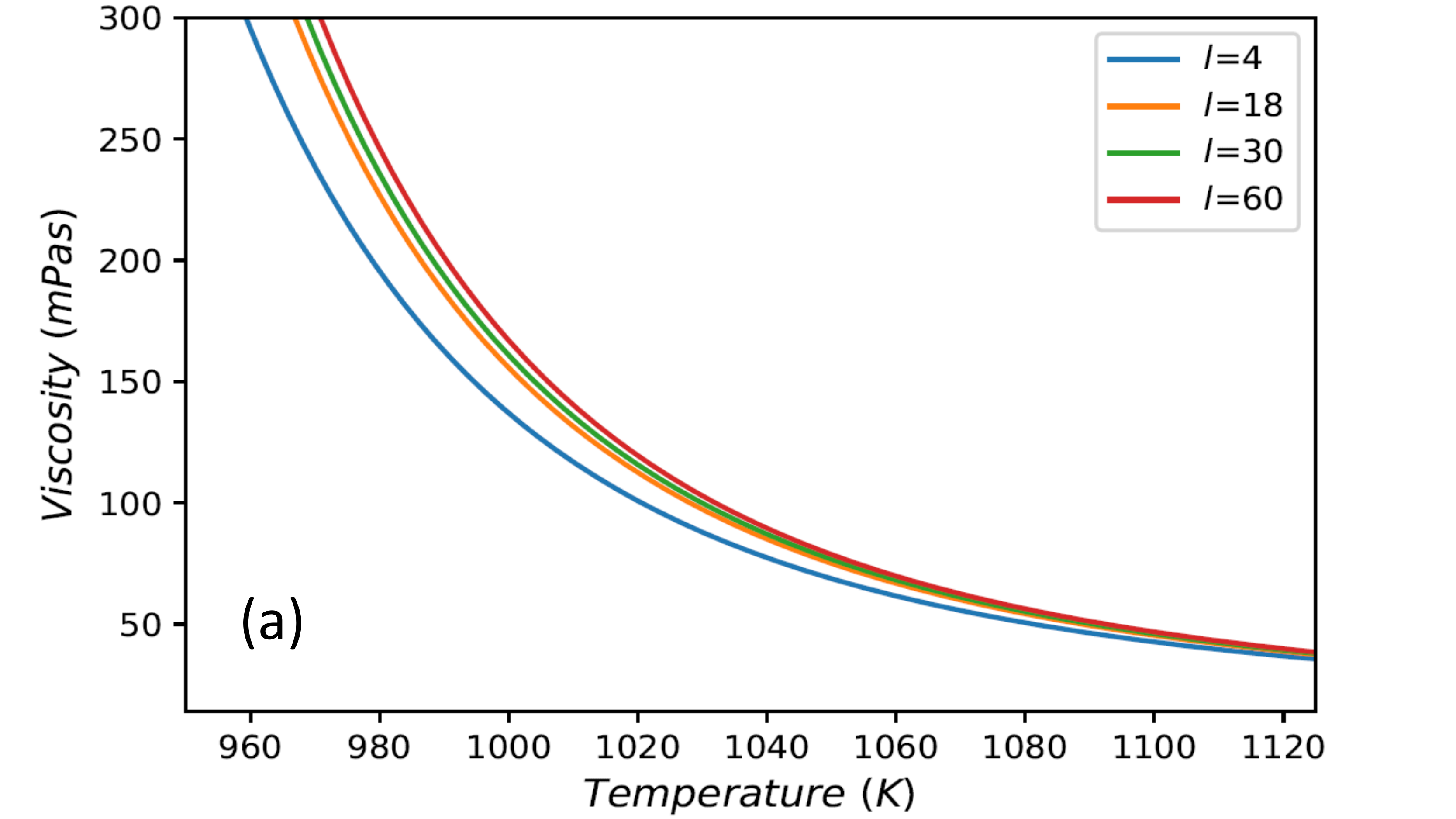}
    \vspace{0.2cm}
    \includegraphics[width=0.85\columnwidth]{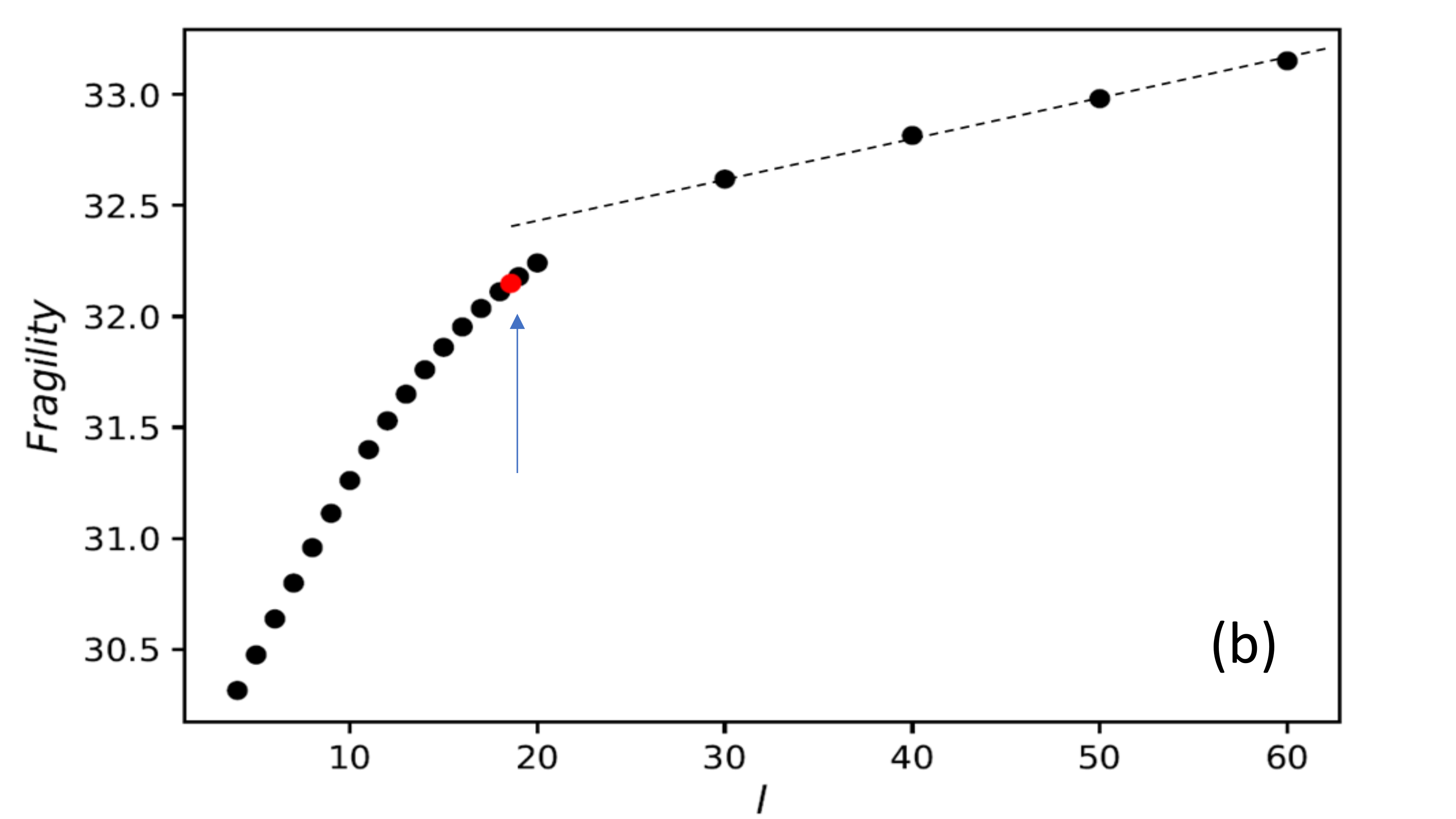}
    \caption{\textbf{(a)}: Viscosity as a function of temperature calculated upon varying the interatomic steepness parameter $l$ around the value $l=18.77$ (red symbol) used in the fitting of Fig. 1 (a). From left to right $l =4, 18, 30, 60$. \textbf{(b)}: Fragility $m$ calculated upon varying the interatomic repulsion steepness parameter $l$. Here the data point with $l \approx 20$ corresponding to Cu$_{50}$Zr$_{50}$ is highlighted by the arrow.}
    \label{fig:tre}
\end{figure}

The fragility of a glass-forming liquid is given as the slope of the viscosity at glass transition temperature, that is: $m=\left.\left(\frac{\partial\log_{10}(\eta/\eta_0)}{\partial(T_g/T)}\right)\right| _{T=T_g}$ \cite{angell2000relaxation}.
Upon applying the definition to our viscosity formula Eq.~\eqref{viscosity}, we obtain the following expression:
\begin{equation}\label{secondfragility}
m(l)=\frac{V_c\alpha_TT_g}{\ln(10)}\rho(T_g)\left(1+\frac{4\pi}{15}\rho(T_g)I(l)\right),
\end{equation}
which can be evaluated by computing $I$ for different values of $l$, by keeping all the other model parameters the same as in the fitting of Fig. 2. 

The fragility evaluated according to Eq.~(7) is plotted in Fig. 4 (b) as a function of the interatomic repulsion steepness parameter $l$. 
It is evident that the fragility $m$ increases monotonically with the repulsion steepness $l$, or, in other words, the fragility $m$ is lower (the liquid is stronger) with softer repulsion steepness. This fact reaffirms the conclusions of Ref.~\cite{krausser2015interatomic} that "softer atoms make stronger liquids", with a surprising robustness of this law across different materials, from colloids~\cite{Weitz} to metals.

It is important to note that the estimate via Eq.~(7) is not \textit{quantitatively} predictive because we used a viscosity fitting calibrated at significantly higher $T$ than $T_g$ where, by definition, the fragility should be evaluated (recall $T_g = 700$K for Cu$_{50}$Zr$_{50}$). 
However, Eq.(7) is valid and reliable as a prediction of the \textit{qualitative} trend for $m$ as a function of $l$. This is because the only parameter in Eq.~(7) which depends on $l$ is the factor $I$, while all other parameters are independent of the interatomic repulsion steepness and act mainly as scale factors. This includes $\alpha_T$ which depends on the attractive part of the potential but not so much on the repulsive part~\cite{kittel1996introduction}.

It is to be noted that the fragility $m$ in Fig. 4 is predicted to go through a crossover from a rapidly varying increasing trend for $l<20$ to a linear trend for $l>20$. The latter linear trend perfectly recovers what has been found with the KSZ equation in ~\cite{krausser2015interatomic}, indeed in a range of larger $m$ values.
Instead, the crossover from a quickly rising initial trend of $m$ with repulsion $l$ into the (more slowly growing) linear $m$ vs $l$ regime is a new prediction of the present work. 

\section{Conclusions}
In summary, a microscopic model of the viscosity and fragile-strong behavior of liquid metals in the supercooled regime has been developed, based on combining the shoving model with the Zwanzig-Mountain (ZM) formula for the high-frequency shear modulus, $G_\infty$ of liquids. 
The model has been evaluated using MD simulations data for the $g(r)$ of the Cu$_{50}$Zr$_{50}$ alloy, from which the interatomic potential of mean force, $\phi(r)$ has been determined upon analytically parameterizing the $g(r)$ data. These input allowed us to evaluate the $G_\infty$ with the ZM formula, and in turn the viscosity as a function of $T$. This led to a semi-analytical formula for $\eta(T)$ which provided an excellent three-parameter fit of experimental viscosity data for the Cu$_{50}$Zr$_{50}$ alloy from the literature~\cite{lagogianni2016unifying}. Compared to other popular three-parameter models, such as VFT and the Mauro equation~\cite{Mauro}, this expression, Eq.~(5), features only microscopic parameters and provides direct access to atomic-scale structural and interaction quantities, such as $g(r)$ and $\phi(r)$. 

The analytical parameterization of the $g(r)$ led to the possibility of studying the slope of viscosity as a function of $T$ and hence the fragile-strong behavior of the liquid, in terms of microscopic interaction potential parameters. It has been shown that a crucial parameter which controls the fragility is the interatomic potential repulsion steepness $l$ (related to Born-Mayer repulsion and to the $\lambda$  parameter of the KSZ equation~\cite{krausser2015interatomic}). It has been demonstrated that the fragility $m$ of an atomic liquid is a monotonically increasing function of the potential repulsion steepness $l$.
This result is independent of the values of the other parameters entering the fragility formula Eq. (7), and depends exclusively on the integral $I$ in the ZM formula, which gives a direct insight into the microscopic explanation for this phenomenon. 
This analysis thus reaffirms that "soft atoms make strong liquids", in full agreement with previous claims on vastly different materials such as metal alloys~\cite{krausser2015interatomic} and colloids~\cite{Weitz,Gnan2019}.

Finally, we also note that the above double-exponential form for $\eta(T)$ Eqs.(5)-(6) has the potential to effectively describe the non-Arrhenius to Arrhenius crossover that has been observed in supercooled liquids~\cite{Stickel1,Stickel2,Roland,Novikov,Schoenhals,Trachenko_Stickel,Zhang}, since a crossover to a single-exponential Arrhenius form (Eq. (1)) is predicted to occur when the thermal expansion coefficient $\alpha_{T}$ is very low. In future work, this consideration may be the starting point to rationalize the extreme variability of the non-Arrhenius to Arrhenius crossover with material chemistries and bonding in terms of the bonding anharmonicity. \\

\begin{acknowledgments}
A.Z. acknowledges financial support from US Army Research Laboratory and US Army Research Office through contract nr. W911NF-19-2-0055.
\end{acknowledgments}

\bibliography{bibliografia} 

\end{document}